\documentclass{elsart}

\usepackage{amsmath,amssymb}
\usepackage{graphicx}

\begin{document}
\begin{frontmatter}
\title{Instanton and Higher-Loop Perturbative Contributions to the 
QCD Sum-Rule Analysis of Pseudoscalar Gluonium}
\author[sask,itp]{Ailin Zhang}
\author[sask]{T.G.\ Steele}
\address[sask]{Department of Physics and Engineering Physics, University of
Saskatchewan, Saskatoon, SK, S7N 5E2, Canada}
\address[itp]{Institute of Theoretical Physics, P.O.\ Box 2735, Beijing,
100080,
P.R.\ China}

\begin{abstract}
Instanton effects and three-loop perturbative contributions are incorporated into 
QCD sum-rule analyses of pseudoscalar ($J^{PC}=0^{-+}$) gluonium.
Gaussian sum-rules are shown to be superior to Laplace sum-rules 
in optimized predictions for pseudoscalar gluonium states in the presence of instanton
contributions.  The Gaussian sum-rule analysis yields a pseudoscalar
mass of  $(2.65\pm 0.33)\, {\rm GeV}$ and width  bounded by
$\Gamma<530\,{\rm MeV}$. The Laplace sum-rules provide corroborating
evidence in support of the $\approx 2.7\, {\rm GeV}$ mass scale.
\end{abstract}


\end{frontmatter}

\section{Introduction}
The gluon self-interaction in QCD suggests the existence of  bound
gluonic states known as gluonium or glueballs.
The properties  of gluonium states have been studied within a wide variety of theoretical 
methods including lattice simulations \cite{lattice,UKQCD}, 
QCD sum-rules  
\cite{nsvz,scalar_sr,ps_sr,bagan_glue,mann,forkel,harnett,two_res_glue,shuryak}, 
and other phenomenological models \cite{model}.
From the experimental viewpoint, a number of scalar and pseudoscalar isoscalar 
states exist which are potential gluonium candidates \cite{pdg}, but a
consensus on the nature of these states has not been achieved. Similarly, theoretical investigations of the 
gluonium mass spectrum continue to be refined.

The initial QCD sum-rule 
analyses of scalar gluonium \cite{scalar_sr} and pseudoscalar gluonium \cite{ps_sr}
have been extended  to include higher-loop contributions of QCD condensates
 and higher-loop perturbative effects \cite{bagan_glue,mann}.  
The non-leading condensate corrections arising from the gluon condensate 
$\langle\alpha G^2\rangle$ \cite{bagan_steele} are particularly significant since they
provide the leading contribution in many sum-rules.

It is also known that (direct) instantons  contribute to the
pseudoscalar and scalar channels \cite{basic_instanton}.  These contributions have 
a significant effect on the scalar gluonium Laplace 
\cite{nsvz,forkel,harnett,shuryak} and Gaussian sum-rules \cite{gauss}.  
In particular, the instanton effects improve the self-consistency of the  lowest-weighted
Laplace sum-rule containing the low-energy theorem and the higher-weight sum-rules 
which are independent of the low-energy theorem \cite{harnett,gauss}.  However, the 
role of instantons in the Laplace sum-rules for pseudoscalar gluonium have not 
been studied in similar detail.  The self-dual properties  of the instanton implies 
that the leading instanton effects  in the scalar and pseudoscalar 
gluonic channels differ only by an overall sign.  This presents the
interesting possibility that instantons are responsible for the large
mass splitting between scalar and pseudoscalar gluonium states
observed in lattice simulations \cite{lattice}.

In this paper, instanton and higher-loop perturbative effects are
incorporated into the Laplace and Gaussian QCD sum-rules for
pseudoscalar gluonium.  In Section \ref{lap_anal_sec} it is demonstrated
that the Laplace sum-rules fail to yield an optimized mass prediction for the
pseudoscalar gluonium state.  This shortcoming is shown to be overcome
by the Gaussian sum-rule approach presented in Section
\ref{gauss_sec}.  The full phenomenological analysis of the Gaussian
sum-rules for pseudoscalar gluonium is presented in Section \ref{gauss_results_sec}.

\section{QCD Laplace Sum-Rules for Pseudoscalar Gluonium}\label{lap_sec}
Pseudoscalar gluonium is studied through the two-point correlation function of the 
renormalization-group invariant currents
 $j(x)=\alpha_s G^a_{\mu\nu}\tilde 
G^a_{\mu\nu}=\alpha_s G^a_{\mu\nu}\frac{1}{2}\epsilon_{\mu\nu\alpha\beta}G^a_{\alpha\beta}$ 
\begin{equation}
\Pi\left(Q^2\right)=i\int \mathrm{d}^4x\,\mathrm{e}^{iq\cdot x}\langle
0|T[J(x)J(0)]|0\rangle, ~~Q^2=-q^2>0
\quad .
\label{corr_fn}
\end{equation}
The correlation function (\ref{corr_fn}) satisfies the dispersion relation
\begin{equation}
\Pi\left(Q^2\right)=\Pi(0)+Q^2\Pi'(0)+\frac{1}{2}Q^4\Pi''(0)
-Q^6\frac{1}{\pi}\int\limits_{t_0}^\infty
\mathrm{d}t\,\frac{ \rho(t)}{t^3\left(t+Q^2\right)} 
\label{disp_rel}
\end{equation}
relating the QCD prediction $\Pi\left(Q^2\right)$ to the 
 hadronic spectral function  $\rho(t)$ appropriate to pseudoscalar gluonic states
 above the physical threshold $t_0$ . The Laplace sum-rules resulting from 
(\ref{disp_rel}) are
\begin{equation}
{\mathcal{L}}_{k}(\tau)=\frac{1}{\pi}\int\limits_{t_0}^\infty t^k
\mathrm{e}^{-t\tau}\rho(t)\,\d t
\quad ,\quad k=0,1,2\ldots\quad ,
\label{lap_gen_k}
\end{equation}
where the theoretically-determined quantity ${\mathcal{L}}_{k}(\tau)$ is
obtained by applying the Borel-transform operator $\hat B$
\begin{equation}
\hat B\equiv 
\lim_{\stackrel{N,~Q^2\rightarrow \infty}{N/Q^2\equiv \tau}}
\frac{\left(-Q^2\right)^N}{\Gamma(N)}\left(\frac{\mathrm{d}}{\mathrm{d}Q^2}\right)^N
\label{borel}
\end{equation}  
to the appropriately-weighted correlation function
\begin{equation}
{\mathcal{L}}_k(\tau)\equiv\frac{1}{\tau}\hat B\left[\left(-1\right)^k Q^{2k}\Pi\left(Q^2\right)\right] . 
\label{laplace}
\end{equation}

In the resonance(s) plus continuum model \cite{SR},  hadronic physics
is locally dual to the QCD 
prediction for energies above the continuum threshold $t=s_0$
\begin{equation}
\rho(t)=\rho^{had}(t)+\theta\left(t-s_0\right){\rm Im}\Pi^{QCD}(t)\quad .
\label{res_plus_cont}
\end{equation} 
The QCD continuum contribution
\begin{equation}
c_k\left(\tau,s_0\right)=\frac{1}{\pi}\int\limits_{s_0}^\infty
t^k \mathrm{e}^{-t\tau}{\rm Im}\Pi^{QCD}(t)\,\mathrm{d}t
\label{continuum}
\end{equation}
is thus combined with the quantity ${\mathcal{L}}_k(\tau)$ because both are QCD predictions, 
resulting in the following Laplace sum-rules relating QCD to hadronic physics phenomenology:
\begin{gather}
R_{k}\left(\tau,s_0\right)=\frac{1}{\pi}\int\limits_{t_0}^{s_0}
t^k \mathrm{e}^{-t\tau}\rho^{had}(t)\,\mathrm{d}t
\label{lap_k}\\
R_k\left(\tau,s_0\right)= {\mathcal{L}}_k\left(\tau\right)-c_k\left(\tau,s_0\right)
\quad .
\end{gather}

\begin{figure}[hbt]
\centering
\includegraphics[scale=0.3,angle=270]{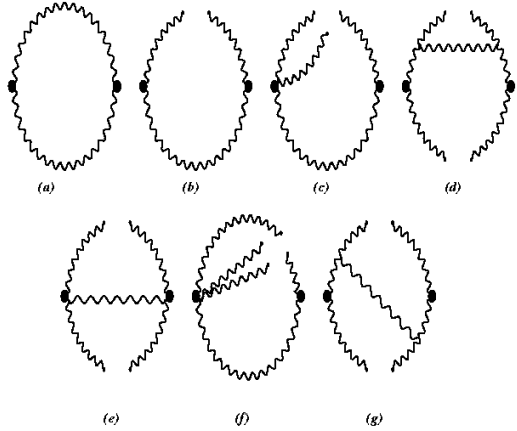}
\caption{
Feynman diagrams used to calculate the leading order perturbative and condensate contributions to
(\protect\ref{correlator}).  Diagram (a) corresponds to the perturbative contribution, diagram (b)
is used to calculate the coefficient of $\langle\alpha G^2\rangle$, diagrams (c) and (d) are 
used to determine the contribution of $\langle g G^3\rangle$, and diagrams (e)--(g) are needed
for the dimension-eight gluonic condensates. Additional diagrams with identical topologies 
have been omitted from the Figure.
}
\label{feyn_fig}
\end{figure}

The field-theoretical (QCD) calculation of $\Pi\left(Q^2\right)$ contains perturbative
(logarithmic) corrections known up to three-loop order
in the chiral limit of $n_f=3$ massless quarks in the $\overline{{\rm MS}}$ scheme \cite{chetyrkin}, 
QCD condensate contributions \cite{mann,condensate_ref}, and direct instantons \cite{nsvz,inst_K2}
\begin{equation}
\begin{split}
\Pi\left(Q^2\right)=&Q^4\log\left(\frac{Q^2}{\nu^2}\right)\left[
a_0+a_1\log\left(\frac{Q^2}{\nu^2}\right)+a_2\log^2\left(\frac{Q^2}{\nu^2}\right)
\right]\\
&+\left[ b_0+b_1\log\left(\frac{Q^2}{\nu^2}\right)\right]\left\langle \alpha G^2\right\rangle
+c_0\frac{1}{Q^2}\left\langle gG^3\right\rangle+d_0\frac{1}{Q^4}
\left\langle {\mathcal{O}}_8\right\rangle\\
&-32\pi^2Q^4\int \rho^4 \left[K_2\left(\rho\sqrt{Q^2}\right)\right]^2  \mathrm{d}n(\rho)
\end{split}
\label{correlator}
\end{equation}   
where
\begin{gather}
a_0=-2\left(\frac{\alpha}{\pi}\right)^2\left[1+20.75\frac{\alpha}{\pi}+
305.95\left( \frac{\alpha}{\pi}\right)^2\right]
\\
a_1=2\left(\frac{\alpha}{\pi}\right)^3\left[ \frac{9}{4}+72.531\frac{\alpha}{\pi}\right]
~ , ~ 
a_2=-10.1250\left(\frac{\alpha}{\pi}\right)^4
\\
b_0=4\pi\frac{\alpha}{\pi}
\quad ,\quad b_1=9\pi\left(\frac{\alpha}{\pi}\right)^2
\quad ,\quad
c_0=-8\pi^2\left(\frac{\alpha}{\pi}\right)^2\quad ,\quad d_0=8\pi^2\frac{\alpha}{\pi}
\\
\left\langle\alpha G^2\right\rangle=\left\langle\alpha G^a_{\mu\nu}G^{a\,\mu\nu}\right\rangle~,~
\left\langle gG^3\right\rangle=\left\langle g f_{abc}G^a_{\mu\nu}G^b_{\nu\rho}G^c_{\rho\mu}\right\rangle
\\
\left\langle {\mathcal{ O}}_8\right\rangle=\left\langle\left(\alpha f_{abc}G^a_{\mu\rho}G^b_{\nu\rho}
\right)^2\right\rangle
+10\left\langle\left(\alpha f_{abc}G^a_{\mu\nu}G^b_{\rho\lambda}\right)^2\right\rangle
\end{gather}
and $K_2(x)$ is the modified Bessel function of the second kind in the conventions of \cite{abram}.
Divergent polynomials corresponding to subtraction constants in 
the dispersion relation have been 
ignored because they are annihilated by the Borel transform, and hence do not contribute to the Laplace 
sum-rules.  The Feynman diagrams used to calculate the leading order perturbative and
condensate contributions  are illustrated in 
Figure \ref{feyn_fig}.\footnote{See Ref.\ \protect\cite{ope} for an outline of several techniques for calculating 
the condensate contributions through the operator-product expansion
and for a proof that such methods are equivalent.}

Although 
the coefficient $b_0$ of the gluon condensate has only been calculated to leading order for the 
pseudoscalar correlation function, the one-loop coefficient $b_1$ is entirely determined by the leading 
order $b_0$ combined with the renormalization group \cite{mann}.  As will be seen below, $b_0$ does not 
enter the Laplace sum-rules 
$\mathcal{L}_k$  for $k\ge 0$, and hence it is crucial to include $b_1$ to obtain the leading effects 
from $\langle \alpha G^2\rangle$.
Finally, it should be noted that the three-loop perturbative calculation 
\cite{chetyrkin} uncovered an error in the two-loop expression  \cite{kataev} 
used in the
sum-rule analysis \cite{mann}, providing further motivation for revisiting this sum-rule analysis of 
pseudoscalar gluonium. 

The instanton contribution in (\ref{correlator}), representing
non-interacting instantons of size $\rho$ with subsequent integration over the instanton density $n(\rho)$, 
is expected to be a reasonable approximation since multi-instanton effects have been found to be
controllable \cite{schaefer_shuryak}.  The self-dual nature of the instanton
implies that the instanton contributions to the pseudoscalar and
scalar gluonium correlators only differ by an overall sign.

The Laplace sum-rules can be constructed from (\ref{correlator}) by using the results contained in
\cite{harnett}
\begin{equation}
\begin{split}
\mathcal{L}_{0}(\tau) =& \frac{1}{\tau^3}\left[-2a_0+a_1\left(-6+4\gamma_E\right)
  +a_2\left(\pi^2-6+18\gamma_E-6\gamma_E^2\right)\right]
\\
  &-\frac{b_1}{\tau}\left\langle \alpha G^2\right\rangle
  +c_0\left\langle gG^3\right\rangle
  +d_0\tau\left\langle\mathcal{O}_8\right\rangle
  \\
  &  -128\pi^2\int\mathrm{d}n(\rho)\ \frac{a^4\mathrm{e}^{-a}}{\rho^2}
       \left[\; 2aK_0(a)+(1+2a)K_1(a)\;\right]
\end{split}
\label{L0}
\end{equation}
\begin{gather}
\begin{split}
\mathcal{L}_{1}(\tau) =& \frac{1}{\tau^4}\left[-6a_0+a_1\left(-22+12\gamma_E\right)
  +a_2\left(3\pi^2-36+66\gamma_E-18\gamma_E^2\right)\right]
\\
 &-\frac{b_1}{\tau^2}\left\langle \alpha G^2\right\rangle
  -d_0\left\langle\mathcal{O}_8\right\rangle
\\ 
  &  -256\pi^2 \int\mathrm{d}n(\rho)\ \frac{a^5\mathrm{e}^{-a}}{\rho^4}
       \left[\; (9-4a)aK_0(a) + (3+7a-4a^2)K_1(a)\;\right]
\end{split}
\label{L1}\\
\begin{split}
c_k\left(\tau,s_0\right)=&
\int\limits_{s_0}^{\infty}
t^{k+2}\mathrm{e}^{-t\tau} \left[-a_0-2a_1\log\left(t\tau\right)
+a_2\left(\pi^2-3\log^2\left(t\tau\right)\right)\right]\,\mathrm{d}t\\
&-b_1\left\langle \alpha G^2\right\rangle\int\limits_{s_0}^\infty
t^k\e^{-t\tau}\d t
\\
& +16\pi^3\int\mathrm{d}n(\rho) \rho^4\int\limits_{s_0}^\infty t^{k+2}J_2\left(\rho\sqrt{t}\right)
  Y_2\left(\rho\sqrt{t}\right)\mathrm{e}^{-t\tau} \,\mathrm{d}t\quad ,
\end{split}
\label{c_k}
\end{gather}
where $a=\rho^2/2\tau$ and it should be noted
 that the one-loop
 term provides  the leading  $\left\langle\alpha G^2\right\rangle$
 contribution to the Laplace sum-rules. 
  In obtaining these expressions,  
renormalization-group improvement has been implemented by setting $\nu^2=1/\tau$ 
after calculating Borel transforms \cite{RG_improve}, so that in the perturbative corrections, $\alpha$ 
is implicitly the three-loop $n_f=3$ $\overline{{\rm MS}}$  running coupling
\begin{gather}
\frac{\alpha_s(\nu)}{\pi}= \frac{1}{\beta_0 L}-\frac{\bar\beta_1\log L}{\left(\beta_0L\right)^2}+
\frac{1}{\left(\beta_0 L\right)^3}\left[
\bar\beta_1^2\left(\log^2 L-\log L -1\right) +\bar\beta_2\right]
\label{alpha_hl}\\
L=\log\left(\frac{\nu^2}{\Lambda^2}\right)~ ,~ \bar\beta_i=\frac{\beta_i}{\beta_0}
~ ,
~
\beta_0=\frac{9}{4}~ ,~ \beta_1=4~ ,~ \beta_2=\frac{3863}{384}~ ,
\end{gather}
with 
 $\Lambda_{\overline{MS}}\approx 300\,{\rm MeV}$ for three active flavours,
consistent with current estimates of $\alpha_s(M_\tau)$ \cite{pdg}.

\section{Laplace Sum-Rule Analysis of Pseudoscalar Gluonium}\label{lap_anal_sec}
The instanton liquid model \cite{ins_liquid}
\begin{equation}
\d n(\rho)=n_c\delta\left(\rho-\rho_c\right)\d\rho\quad ;\quad n_c=8\times 10^{-4}\,{\rm GeV^4}\quad ,\quad
\rho_c=\frac{1}{600 \,{\rm MeV}}
\label{ins_liquid_params}
\end{equation}
will be used in the Laplace sum-rule analysis, along with the  value for the the 
dimension-eight condensates obtained from vacuum saturation and the
heavy-quark expansion \cite{nsvz,vacuum_saturation}
\begin{equation}
\left\langle \mathcal{O}_8\right\rangle=
\frac{15}{16}\left(\left\langle \alpha G^2\right\rangle\right)^2
\label{vac_sat}
\end{equation}
and the instanton estimate of the dimension-six gluon condensate \cite{nsvz,SR}
\begin{equation}
\left\langle gG^3\right\rangle
=\left(0.27\,{\rm GeV^2}\right)\left\langle \alpha G^2\right\rangle\quad .
\end{equation}
The dimension-six and dimension-eight condensates are thus related to the
gluon condensate given by  the
determination \cite{GG_ref}
\begin{equation}
\left\langle \alpha G^2\right\rangle=(0.07\pm 0.01) \,{\rm GeV^4}\quad .
\label{GG_cond}
\end{equation}

In the narrow resonance(s) model
\begin{equation}
\rho^{had}(t)=\pi \sum F^2M^4\delta\left(t-M^2\right)\quad,
\label{narrow}
\end{equation}
resonance masses are signaled by exponential decay of the sum-rules
\begin{equation}
R_k\left(\tau,s_0\right)=\sum  F^2M^{4+2k}\exp{\left(-M^2\tau\right)}\quad .
\label{nar_res_SR}
\end{equation}

\begin{figure}[hbt]
\centering
\includegraphics[scale=0.45]{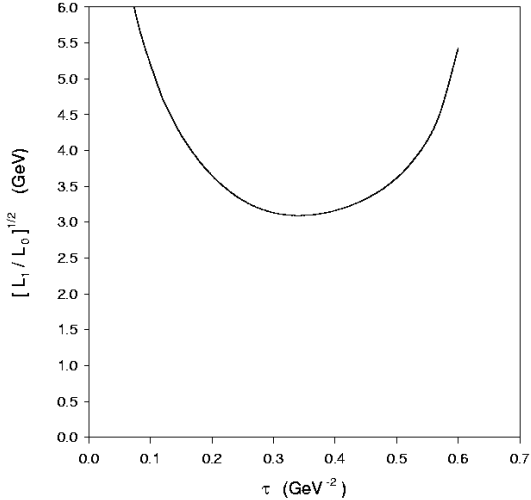}
\caption{
The ratio $\sqrt{\mathcal{L}_1/\mathcal{L}_0}$ as a function of $\tau$ for central values of the QCD
parameters.   
}
\label{bound_fig}
\end{figure}

A bound on the mass $M$ of the lightest state can then be obtained from the ratio
\begin{equation}
\frac{\mathcal{L}_1(\tau)}{\mathcal{L}_0(\tau)}
=\lim_{s_0\to\infty}
\frac{R_1\left(\tau,s_0\right)}{R_0\left(\tau,s_0\right)}\ge M^2 \quad .
\label{bounds}
\end{equation}
This bound is quite robust since it does not depend on the QCD
continuum approximation and does not require dominance from the
lightest state (see {\it e.g.} \cite{harnett}). Figure \ref{bound_fig}
displays the ratio (\ref{bounds}), resulting in a bound on the
lightest pseudoscalar gluonium state of
$M<3.1\,{\rm GeV}$.

\begin{figure}[hbt]
\centering
\includegraphics[scale=0.45]{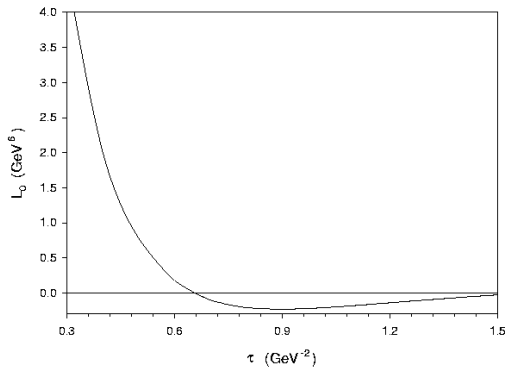}
\caption{
The sum-rule
$\mathcal{L}_0\left(\tau\right)=\lim_{s_0\to\infty}R_0\left(\tau,s_0\right)$
as a function of $\tau$.
Central values of the QCD
parameters have been used.   
}
\label{L0_fig}
\end{figure}

\begin{figure}[hbt]
\centering
\includegraphics[scale=0.45]{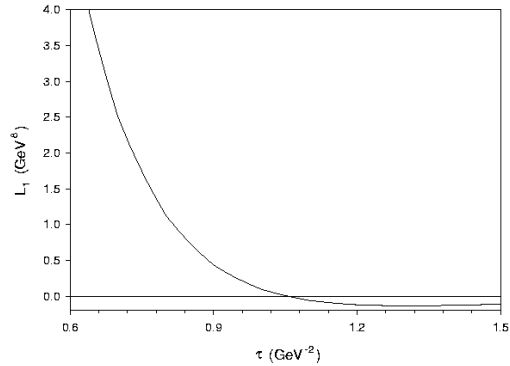}
\caption{
The sum-rule
$\mathcal{L}_1\left(\tau\right)=\lim_{s_0\to\infty}R_1\left(\tau,s_0\right)$
as a function of $\tau$.
Central values of the QCD
parameters have been used.   
}
\label{L1_fig}
\end{figure}

As illustrated in Figures \ref{L0_fig} and \ref{L1_fig},
the rise past the minimum in Figure \ref{bound_fig} is actually a
manifestation of a zero which occurs in $\mathcal{L}_0(\tau)$ prior to
the first zero of  $\mathcal{L}_1(\tau)$, resulting in a singularity
of the ratio (\ref{bounds}).  Negative values of $\mathcal{L}_k(\tau)$
are inconsistent with (\ref{laplace}) and  positivity of the spectral
function $\rho(t)$. This unphysical behaviour of the sum-rules can be traced
to the instanton contributions in the low-energy (large $\tau$)
regime.

\begin{figure}[hbt]
\centering
\includegraphics[scale=0.45]{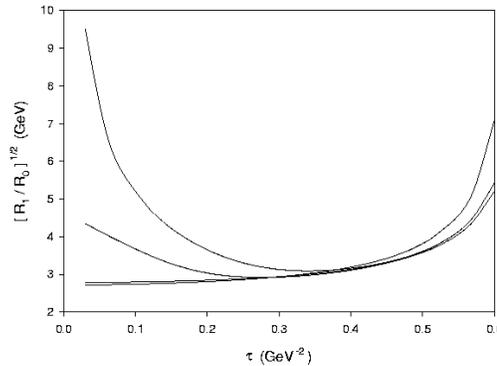}
\caption{
Sum-rule ratios
$\sqrt{R_1\left(\tau,s_0\right)/R_0\left(\tau,s_0\right)}$ ranging from
$s_0=\infty$ ({\it i.e.}  $\sqrt{\mathcal{L}_1/\mathcal{L}_0}$) in the
top
curve to $s_0=8\,{\rm GeV^2}$ in the bottom curve.
Central values of the QCD
parameters have been used.   
}
\label{ratios_fig}
\end{figure}

The unphysical low-energy behaviour of the sum-rules does
not necessarily invalidate the bounds obtained from Figure \ref{bound_fig}
because the minimum occurs in the perturbative (high-energy, low
$\tau$) regime, but it does present difficulties in the extraction of
an optimized mass estimate.  Within the single narrow resonance model,   
(\ref{nar_res_SR}) results in 
\begin{equation}
\frac{R_1\left(\tau,s_0\right)}{R_0\left(\tau,s_0\right)}=M^2\quad.
\label{mass_rat}
\end{equation}
In a typical sum-rule analysis \cite{SR}, one decreases $s_0$ until a stable
({\it i.e.} nearly $\tau$-independent) ratio (\ref{mass_rat}) is
obtained.  Figure  \ref{ratios_fig} shows that the ratio does stabilize at approximately
$2.7\,{\rm GeV}$, but the small $\tau$ values  associated with this stability
region are problematic because dominance of the lightest resonance
cannot be assured and  uncertainties from the
continuum approximation become significant.   
 However, because (\ref{bounds}) has general validity beyond 
these approximations, the $M<3.1\,{\rm GeV}$ bound remains valid.

Thus instanton effects prevent the extraction of a reliable optimized
Laplace sum-rule mass prediction for pseudoscalar 
gluonium, a difficulty that has also been observed in the configuration-space instanton analysis 
\cite{schaefer_shuryak}.  It is interesting that the resonance signal in each case is exponential decay 
associated with the resonance mass combined with $\tau$ in the Laplace sum-rules (\ref{nar_res_SR}) 
or with Euclideanized time in the configuration-space approach \cite{schaefer_shuryak}.
The next Section will review  the formulation  of Gaussian sum-rules which will be seen to provide a 
fundamentally-different weighting of the hadronic spectral function and QCD contributions, obviating
the problems associated with an optimization of the Laplace sum-rule analysis.
  
\section{Gaussian Sum-Rules for Pseudoscalar Gluonium}\label{gauss_sec}
The simplest Gaussian sum-rule (GSR)  \cite{first_gauss} 
\begin{equation}
G\left( \hat s,\tau\right)=\frac{1}{\pi}\int\limits_{t_0}^\infty\frac{1}{\sqrt{4\pi\tau}}
\exp{\left( -\frac{\left(t-\hat s\right)^2}{4\tau}\right)}\,\rho(t)\,\mathrm{d}t
\quad ,\quad \tau>0
\label{basic_gauss}
\end{equation}
relates the QCD prediction on the left-hand side of (\ref{basic_gauss}) to the hadronic spectral 
function $\rho(t)$ 
smeared over the energy range $\hat s -2\sqrt{\tau}\lesssim t\lesssim\hat s +2\sqrt{\tau}$, representing 
an energy interval for quark-hadron duality.\footnote{The quantity $\tau$ in the GSR should not be
confused with the similar quantity appearing in the Laplace sum-rules. In particular, the
mass dimension is different in each usage.} 
An interesting 
aspect of the GSR is that the duality interval is actually constrained by QCD. A lower bound on this 
duality scale 
$\tau$ necessarily exists because the 
QCD prediction has 
renormalization-group properties that reference running quantities the the energy scale 
$\nu^2=\sqrt{\tau}$ \cite{first_gauss,orl00}.
Thus it is not possible to achieve the formal  $\tau\to 0$ limit where complete knowledge of the spectral 
function could be obtained  via
\begin{equation}
\lim_{\tau\to 0}G\left(\hat s,\tau\right)=\frac{1}{\pi}\rho\left(\hat s\right)\quad ,\quad \hat s>t_0 
\quad .
\label{gauss_limit}
\end{equation}
However, there is no theoretical constraint on the quantity $\hat s$ representing the peak of the Gaussian 
kernel appearing in (\ref{basic_gauss}).  Thus the
$\hat s$ dependence of the QCD prediction $G\left(\hat s,\tau\right)$ probes the behaviour of the smeared 
spectral function, reproducing the essential 
features of the spectral function. In particular, as $\hat s$ passes through $t$ values corresponding to  
resonance peaks, the Gaussian kernel in 
(\ref{basic_gauss}) reaches its maximum value, implying that Gaussian sum-rules weight excited and ground 
states equally.

The QCD prediction for the GSR is obtained from the correlation function through
\begin{equation}\label{srdef}
   G(\hat{s},\tau)\equiv \sqrt{\frac{\tau}{\pi}}\mathcal{B}
   \left\{ \frac{ \Pi\left(-\hat{s}-i\Delta\right)
         -  \Pi\left(-\hat{s}+i\Delta\right) }{i\Delta}
   \right\} \quad ,
\end{equation}
where the Borel transform $\mathcal{B}$ has been redefined as
\begin{equation}
  \mathcal{B}\equiv \lim_{\stackrel{N,\Delta^2\rightarrow\infty}{\Delta^2/N\equiv 4\tau}}
  \frac{(-\Delta^2)^N}{\Gamma(N)}\left( \frac{\mathrm{d}}{\mathrm{d}\Delta^2}\right)^N \quad.
\end{equation}
The GSR (\ref{basic_gauss}) then results from combining (\ref{srdef}) with the dispersion relation
(\ref{disp_rel}) \cite{first_gauss}.

A connection between Gaussian and 
 finite-energy sum-rules can be   established  through the diffusion
equation satisfied by the GSR 
\begin{equation}
\frac{\partial^2 G\left( \hat s,\tau\right)}{\partial\hat s^2}=
\frac{\partial G\left( \hat s,\tau\right)}{\partial \tau}\quad .
\label{diffusion}
\end{equation}
In particular, the resonance(s) plus continuum model (\ref{res_plus_cont}),
when $\rho^{{had}}(t)$ is evolved through the diffusion equation (\ref{diffusion}), only reproduces the 
QCD prediction at large energies ($\tau$ large)
if the resonance and continuum contributions are balanced through the finite-energy sum-rule \cite{first_gauss}
\begin{equation}
F\left(s_0\right)=\frac{1}{\pi}\int\limits_{t_0}^{s_0} \rho^{had}(t)\,\mathrm{d}t  \quad .
\label{basic_fesr}
\end{equation}

Within the resonance(s) plus continuum model (\ref{res_plus_cont}), the continuum contribution to the 
GSR is determined by QCD
\begin{equation}\label{gauss_cont}
   G^{{ cont}} (\hat{s},\tau,s_0) = \frac{1}{\sqrt{4\pi\tau}}   \int_{s_0}^{\infty} 
   \exp \left[ \frac{-(\hat{s}-t)^2}{4\tau} \right]  \frac{1}{\pi} {\rm Im} \Pi^{QCD}(t) \d {t}\quad ,
\end{equation}
and is thus combined with $G_k\left(\hat s,\tau\right)$ to give the total QCD contribution
\begin{equation}
  G^{qcd}\left(\hat{s},\tau,s_0\right) \equiv G\left(\hat{s},\tau\right) -  G^{{cont}} \left(\hat{s},\tau,s_0\right) \quad ,
\label{blah} 
\end{equation}
resulting in the final relation between the QCD and hadronic sides of the GSR.
\begin{equation}
 G^{qcd}\left(\hat{s},\tau,s_0\right)
= \int_{t_0}^{\infty} 
    \exp\left[ \frac{-(\hat{s}-t)^2}{4\tau}\right] \frac{1}{\pi}
    \rho^{had}(t) \d {t}  \quad .
\label{final_gauss}
\end{equation}
Comparison of (\ref{final_gauss}) and (\ref{lap_k}) reveals that the
GSR provides a fundamentally different weighting of the hadronic
spectral function than the Laplace sum-rule.  In particular, the
Laplace sum-rules always emphasize the low-energy region, while
this aspect is controlled in the GSR by the parameter $\hat s$.

The expression for the QCD prediction for the pseudoscalar gluonium GSR can then 
be constructed using the results
of \cite{gauss}.
\begin{equation}
\begin{split}
      G^{qcd}(\hat{s},\tau,s_0) &= - \frac{1}{\sqrt{4\pi\tau}} \int_0^{s_0}\!\! t^2 \d {t}
       \exp\left[ \frac{-(\hat{s}-t)^2}{4\tau}\right]  \Biggl[ (a_0-\pi^2 a_2)
       +2a_1\log\left( \frac{t}{\nu^2} \right)\Biggr. 
\\ 
&\phantom{- \frac{1}{\sqrt{4\pi\tau}} \int_0^{s_0} t^2 \d {t}
       \exp\left[ \frac{-(\hat{s}-t)^2}{4\tau}\right]}\qquad \Biggl.+ 3a_2 \log^2\left( \frac{t}{\nu^2} \right) \Biggr]      
   \\
       &- \frac{1}{\sqrt{4\pi\tau}} b_1\langle \alpha G^2\rangle \int_0^{s_0}
        \exp\left[ \frac{-(\hat{s}-t)^2}{4\tau}\right] \d {t}
\\
       & + \frac{1}{\sqrt{4\pi\tau}} \exp\left( \frac{-\hat{s}^2}{4\tau}\right)
         \left[ c_0 \left\langle {\mathcal{ O}}_6\right\rangle  - \frac{d_0 \hat{s}}{2\tau}
         \left\langle {\mathcal{ O}}_8\right\rangle \right]
   \\
       &+\frac{16\pi^3}{\sqrt{4\pi\tau}} \int\!\!\d {n}(\rho)\rho^4
       \int_0^{s_0} t^2  \exp\left[ \frac{-(\hat{s}-t)^2}{4\tau}\right]  J_2\left(\rho\sqrt{t} \right)
       Y_2\left(\rho\sqrt{t} \right)  \d {t} 
\end{split}
\label{G0}
\end{equation}
As mentioned earlier,
 renormalization-group improvement necessitates the replacement $\nu^2=\sqrt{\tau}$ in (\ref{G0}) 
\cite{first_gauss,orl00}. As with the Laplace sum-rules, the one-loop
 term provides  the leading  $\left\langle\alpha G^2\right\rangle$
 contribution to the GSR. 
For the dimension-six and -eight non-perturbative QCD condensate
 contributions in (\ref{G0}), it is easily seen that the
 non-perturbative corrections are 
exponentially suppressed for large
$\hat s$.   Since $\hat s$ represents the location of the Gaussian peak on the phenomenological side of the sum-rule,
the non-perturbative corrections are most important in the low-energy region, as anticipated by the 
role of  QCD condensates in relation to the vacuum properties of QCD.  This explicit low-energy role of the QCD
condensates clearly exhibited for the Gaussian sum-rules is obscured in the Laplace sum-rules.

Integrating both sides of (\ref{final_gauss}) reveals that
 the overall normalization of the above equation is related to the
finite-energy sum-rule \cite{orl00}
\begin{equation}
  \int\limits_{-\infty}^\infty G^{qcd}(\hat{s}, \tau,s_0) \d {\hat{s}}
  =\frac{1}{\pi}\int\limits_{t_0}^{\infty} \rho^{{had}}(t) \d {t} \quad .
  \label{tom_norm_2}
\end{equation}
Thus the diffusion equation analysis \cite{gauss} relates the
normalization of the GSR to the finite-energy 
sum-rules. Information independent of this 
relation is thus extracted from the {\em normalized} GSR \cite{orl00}
\begin{gather}
  N^{qcd}(\hat{s}, \tau, s_0) \equiv
  \frac{G^{qcd}\left(\hat s, \tau, s_0\right)}{m_{0}\left(\tau, s_0\right)}
  \label{tom_norm_srk}\\
  m_{n}(\tau, s_0)
  = \int\limits_{-\infty}^\infty \hat{s}^n G^{qcd}\left(\hat s,\tau, s_0\right) \d {\hat{s}}
  \quad,\quad n=0,1,2,\ldots\quad,
\label{moments}
\end{gather}
which is related to the hadronic spectral function via  
\begin{equation}\label{ngsr}
   N^{qcd}\left(\hat{s},\tau,s_0\right) = \frac{ \int\limits_{t_0}^{\infty} 
   \exp\left[\frac{-(\hat{s}-t)^2}{4\tau} \right] 
\rho^{{had}}(t)\d {t}}{\sqrt{4\pi\tau}\int\limits_{t_0}^{\infty} 
   \rho^{{had}}(t)\,\d t} \quad.
\end{equation}

\section{Gaussian Sum-Rule Analysis of Pseudoscalar Gluonium}\label{gauss_results_sec}
The techniques for analyzing the GSRs  were initially developed in
\cite{gauss,orl00}.  In the single narrow resonance
model, the normalized GSR (\ref{ngsr}) becomes
\begin{equation}\label{phenom_single}
  N^{qcd}\left(\hat{s},\tau,s_0\right) = \frac{1}{\sqrt{4\pi\tau}}
  \exp\left[ -\frac{(\hat{s}-M^2)^2}{4\tau}\right]
  \quad.
\end{equation}
Deviations from the narrow-width limit are  proportional to
$M^2\Gamma^2/\tau$, so this narrow-width model may 
 be a good numerical approximation.  Phenomenological analysis
of the single narrow resonance model 
proceeds from the observation that, as a function of $\hat s$,  
the phenomenological side of (\ref{phenom_single}) has a maximum value (peak) 
at $\hat s=M^2$ independent of the value of $\tau$.  The value of
$s_0$ is then optimized by minimizing the 
$\tau$ dependence of the $\hat s$ 
peak position of the QCD prediction, and the resulting $\tau$-averaged
$\hat s$ peak position leads to a 
prediction of the resonance mass \cite{orl00}.

For the central values of the QCD parameters, minimizing the $\hat s$
peak motion in the region $2\,{\rm GeV^4}<\tau<4\,{\rm GeV^4}$ results
in $M=2.70\,{\rm GeV}$ for $\sqrt{s_0}=3.3\,{\rm GeV}$, which is  remarkably
similar to the mass scale  resulting from the Laplace sum-rule stability analysis in
Figure \ref{ratios_fig}.\footnote{The  range of $\tau$ is chosen to
  have acceptable convergence of the perturbative series while
  maintaining a resolution consistent with typical hadronic scales. 
   }  
Although an optimized Laplace sum-rule
analysis was not possible, a consistent scenario is emerging from the two
approaches.

\begin{figure}[hbt]
\centering
\includegraphics[scale=0.45]{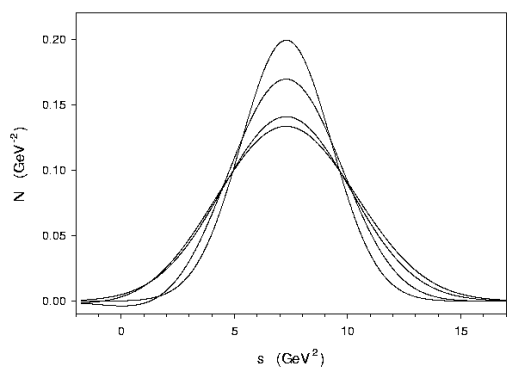}
\caption{
Comparison of the 
QCD and single narrow resonance model contributions to the normalized
GSRs
as a function of $\hat s$
for the optimized parameters $M=2.70\,{\rm GeV}$ and
$\sqrt{s_0}=3.3\,{\rm GeV}$ for $\tau=2\,{\rm GeV^4}$ (upper pair of
curves) and for $\tau=4\,{\rm GeV^4}$ (lower pair of curves).  In each
case, the peak of the theoretical (QCD) prediction lies below that of
the single-resonance model. 
}
\label{single_fig}
\end{figure}

Figure \ref{single_fig} compares the theoretical and phenomenological
sides of the normalized GSR (\ref{ngsr}). It is evident from this
comparison that the single 
narrow resonance model is an inadequate description of the spectral
function predicted by QCD.  Figure \ref{single_fig} also reveals a
region of small $\hat s$ where the theoretical GSR is negative,
inconsistent with a positive spectral function through  (\ref{ngsr}).
This effect can again be traced to the instanton contributions, and
since $\hat s$ corresponds to the peak of the Gaussian kernel, it is
clear that this is a low-energy non-perturbative effect that is safely isolated
from the peak of the theoretical contribution which enters the optimization analysis.

By construction, all the
curves in Figure \ref{single_fig} are normalized to unit area, so the
QCD contributions which underestimate the  peak
 are necessarily broader than those of the single narrow resonance
model.\footnote{This point is explicitly evident in the tails where QCD
  overestimates the single narrow resonance model.}  In particular,  moment combinations  (\ref{moments})
associated with (\ref{phenom_single}) result in
\begin{equation}
\frac{m_2}{m_0}-\left(\frac{m_1}{m_0}\right)^2=2\tau\quad .
\label{width_moments}
\end{equation} 
Thus the narrow-resonance width of $2\tau$ is insufficient to
provide agreement with QCD, indicative of distributed resonance
strength  that can be resolved by the GSR.   

A simple toy model that illustrates the effect of resonance widths in
the broadening of the phenomenological side of the normalized GSR is
a unit-area ``square pulse'' which could describe a broad structureless
feature of the spectral function
\begin{equation}
\frac{1}{\pi}\rho^{{had}}(t)=
\frac{1}{2M\Gamma}\left[
\theta\left(t-M^2+M\Gamma\right)-\theta\left(t-M^2-M\Gamma\right)
\right]\quad ,
\label{square_pulse}
\end{equation}
which leads to the following normalized Gaussian sum-rule \cite{gauss} 
\begin{equation}
 N^{qcd}\left(\hat{s},\tau,s_0\right)
=\frac{1}{4M\Gamma}\left[
{\rm erf}\left(\frac{\hat s-M^2+M\Gamma}{2\sqrt{\tau}}\right)
-{\rm erf}\left(\frac{\hat s-M^2-M\Gamma}{2\sqrt{\tau}}
\right)
\right]\quad .
\label{gauss_sp}
\end{equation}
The second-order moment combinations resulting from this sum-rule are \cite{gauss}
\begin{equation}
\frac{m_2}{m_0}-\left(\frac{m_1}{m_0}\right)^2=2\tau+\frac{1}{3}M^2\Gamma^2\quad ,
\label{more_width_moments}
\end{equation} 
and hence resonance widths broaden the phenomenological side of the
normalized GSR as would be expected intuitively.

For detailed analysis, the following  Gaussian resonance and a skewed-Gaussian
resonance models will be employed since they are numerically
simpler to analyze than a Breit-Wigner  resonance shape.
\begin{gather}
\rho(t)\sim \exp{\left[-\frac{\left(t-M^2\right)^2}{2\gamma^2}\right]}
\label{gauss_res}
\\
\rho(t)\sim t^2\exp{\left[-\frac{\left(t-M^2\right)^2}{2\gamma^2}\right]}
\label{skew_gauss_res}
\end{gather}
The quantity $\gamma$ can be related to a Breit-Wigner width $\Gamma$ by
equating the half-widths, resulting in
$\Gamma=\sqrt{2\log2}\,\gamma/M$.
For the Gaussian model, the resulting normalized GSR is \cite{gauss}
\begin{equation}
  N^{qcd}\left(\hat{s},\tau,s_0\right) =
  \frac{1+\text{erf}\left(\frac{\hat s\gamma^2+2M^2\tau}{2\gamma\sqrt{\tau}\sqrt{\gamma^2+2\tau}}\right)}
  {\sqrt{2\pi}\sqrt{\gamma^2+2\tau}\left[1+\text{erf}\left(\frac{M^2}{\sqrt{2}{\gamma}}\right)\right]}
  \exp\left[-\frac{\left(\hat{s}-M^2\right)^2}{2\left(\gamma^2+2\tau\right)}\right]
\quad .
\label{phenom_gauss}
\end{equation}   
For the skewed Gaussian model, the resulting normalized GSR is
\begin{equation}
\begin{split}
 \frac{ N^{qcd}\left(\hat{s},\tau,s_0\right)}{A} =&
 \exp\left[-\frac{\left(\hat{s}-M^2\right)^2}{2\left(\gamma^2+2\tau\right)}\right] 
\left[
1+\text{erf}\left(\frac{\hat s\gamma^2+2M^2\tau}{2\gamma\sqrt{\tau}\sqrt{\gamma^2+2\tau}}\right)
\right]
\\
&\times
\sqrt{\pi}\left[
\hat s^2\gamma^4+4\hat s\gamma^2\tau M^2+4\tau^2M^4+2\tau\gamma^4+4\tau^2\gamma^2
\right]
\\
&+2\gamma\left(\hat s\gamma^2+2\tau M^2\right)\sqrt{\tau\left(\gamma^2+2\tau\right)}
\exp{\left[
-\frac{\hat s^2\gamma^2+2\tau M^4}{4\tau\gamma^2}
\right]}
\end{split}
\label{skew_gauss_sr}
\end{equation}
where
\begin{equation}
A=\frac{\frac{\gamma}{2\sqrt{\pi}\left(\gamma^2+2\tau\right)^{5/2}}}{M^2\gamma^2\exp{\left(-\frac{M^4}{2\gamma^2}\right)}
+\sqrt{\frac{\pi}{2}}\left(M^4\gamma+\gamma^3\right)\left[1+\text{erf}\left(\frac{M^2}{\gamma\sqrt{2}}\right)\right] }~. 
\end{equation}

The inclusion of resonance widths implies that the $\hat s$ peak
position of the phenomenological side of the sum-rules can develop
$\tau$ dependence, complicating the optimization of $s_0$.  The direct
approach of determining all parameters from the best fit between the
$\hat s$ dependence of the
two sides of (\ref{phenom_gauss}) and (\ref{skew_gauss_sr}) over the
range $2\,{\rm GeV^4}<\tau<4\,{\rm GeV^4}$ is
facilitated by an initial estimate of $s_0$.   An effective estimate
of the optimized $s_0$ is obtained from the (approximate) $\tau$
dependence of the $\hat s$ peak which has the general behaviour \cite{gauss,orl00}
\begin{equation}
\hat s_{peak}\left(\tau,s_0\right)=  a + \frac{b}{\tau} + \frac{c}{\tau^2}\quad .
\label{peak_drift}
\end{equation}
Analysis of how the $\hat s$ peak ``drifts'' with $\tau$ in comparison
with the behaviour (\ref{peak_drift}) 
provides an estimate  of $s_0$ which is found to be surprisingly close to the true
fitted value, facilitating numerical analysis of the multi-dimensional fit.

\begin{figure}[hbt]
\centering
\includegraphics[scale=0.45]{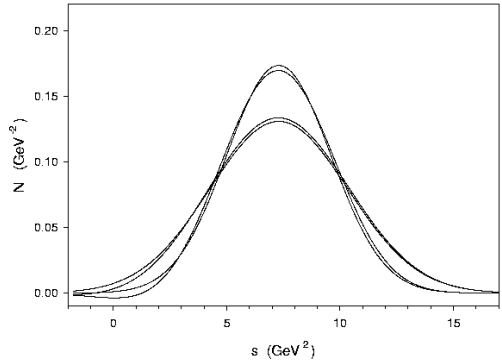}
\caption{
Comparison of the 
QCD and skewed-Gaussian resonance model contributions to the
normalized GSR
as a function of $\hat s$
for the optimized parameters $M=2.6\,{\rm GeV}$, $\Gamma=0.53\,{\rm GeV}$,  and
$\sqrt{s_0}=3.2\,{\rm GeV}$ for $\tau=2\,{\rm GeV^4}$ (upper pair of
curves) and for $\tau=4\,{\rm GeV^4}$ (lower pair of curves).  For the
upper pair of curves 
 the peak of the theoretical (QCD) prediction lies above that of
the single-resonance model, but  the situation is reversed for the
lower pair of curves. 
}
\label{skew_fig}
\end{figure}

For the central values of the QCD parameters, the resulting optimized parameters
are $M=2.7\,{\rm GeV}$, $\Gamma=0.50\,{\rm GeV}$
 in the Gaussian resonance model, and  
 $M=2.6\,{\rm GeV}$, $\Gamma=0.53\,{\rm GeV}$
 in the skewed Gaussian resonance model, with $\sqrt{s_0}=3.2\,{\rm GeV}$ in
 each case.  The agreement between the QCD and phenomenological sides
 of the normalized GSR for the skewed Gaussian model is shown in
 Figure \ref{skew_fig}. Comparison of Figures \ref{single_fig} and
 \ref{skew_fig} (which have identical scales) reveals that inclusion of a resonance width provides a
 significant improvement in the agreement between the QCD and
 phenomenological sides of the sum-rule without a significant change
 in the predicted resonance mass.

The un-skewed Gaussian resonance model results in an agreement between
the QCD and phenomenological sides of the normalized GSR which is
qualitatively and quantitatively indistinguishable from Figure
\ref{skew_fig}.  Unlike the single narrow resonance model where theory
consistently underestimated the phenomenological peak, there is no evident pattern in the disagreement
between  QCD and the skewed-Gaussian model. 
It is possible that the slight disagreement that
remains between QCD and the phenomenological model can be explained by
the low-energy region where $G^{qcd}$ is negative, since this 
 reduces the quantity $m_0$ for $\tau=2\,{\rm GeV^4}$, and
hence raises the value of $N^{qcd}$. Thus there is no clear motivation for
considering more complex resonance models.

The effect of varying the QCD instanton parameters by 15\% and the
gluon condensate within the range (\ref{GG_cond})  in both the Gaussian
and skewed-Gaussian models provides a final
determination of the mass and width of pseudoscalar gluonium
\begin{equation} 
M=(2.65\pm 0.33)\, {\rm GeV}\quad ,\quad
\Gamma<540\,{\rm MeV}\quad .
\label{final_numbers}
\end{equation}
An interesting aspect of this analysis is that smaller instanton sizes
are essentially consistent with a narrow resonance model. 

\section{Conclusions}
Instanton and higher-loop perturbative effects have been incorporated into
the Laplace and Gaussian sum-rules for pseudoscalar gluonium.
Although the Gaussian sum-rules are found to be  superior to the
Laplace sum-rules in obtaining an optimized prediction of the
pseudoscalar gluonium properties, the resulting mass scales in each analysis are
remarkably consistent,   providing corroborating evidence in support
of our final mass prediction of $M\approx 2.7\,{\rm GeV}$.  
Our mass estimate is consistent with the benchmark lattice scalar
glueball mass of approximately $1.6\,{\rm GeV}$ and a 
pseudoscalar to scalar mass ratio of approximately 1.5 \cite{lattice}.

The analysis presented in this paper has not pursued the interesting possibility of mixing
between gluonium and quark mesons \cite{mixing}, although in principle
any hadronic state that overlaps with the pseudoscalar gluonium
interpolating field would be probed by the correlator (\ref{corr_fn})
which  underlies  our entire analysis.  

This possibility of other 
resonances coupled to the pseudoscalar gluonic current was explored 
through an extension of the  Section \ref{gauss_results_sec}
analysis to two narrow resonances tends to lead to
nearly-degenerate states near $2.7\,{\rm GeV}$, suggesting that the
$2.7\,{\rm GeV}$ state is more strongly  coupled to the
pseudoscalar gluonic operator than other pseudoscalar states such as
the $\eta'$. This result is not unexpected, since the coupling
of the $2.7\,{\rm GeV}$ state $\vert P\rangle$ to the pseudoscalar gluonic current 
 extracted  from the peak value of $G^{qcd}\left(\hat s,\tau,s_0\right)$
with the optimized value of $s_0$ is
\begin{equation}
\left|\langle O\left|\alpha G\tilde G\right|P\rangle\right|\approx
20\,{\rm GeV^3}\quad .
\label{P_coupling}
\end{equation}
By comparison, the couplings of $\eta,\eta'$ to the pseudoscalar
gluonic operators are \cite{eta_couplings}
\begin{equation}
\left|\langle O\left|\alpha G\tilde G\right|\eta'\rangle\right|\approx
0.9\,{\rm GeV^3}~ ,~
\left|\langle O\left|\alpha G\tilde G\right|\eta\rangle\right|\approx
0.1\,{\rm GeV^3}~,
\label{eta_coup}
\end{equation}
and hence the relative strength of the $\vert\eta'\rangle$ compared
with $\vert P\rangle$ obtained from the
ratios of the squares of (\ref{P_coupling}) and (\ref{eta_coup}) is
approximately 0.2\%, and the relative strength of the $\vert\eta\rangle$ is 
approximately 0.02\%.  This result is upheld by a fit of the
relative strength of the $\eta'$ and a Gaussian resonance to the
normalized GSR results in a relative strength of less than 1\%.
 Thus the  $2.7\,{\rm
  GeV}$ state resulting from our analysis appears to be  dominantly-coupled
to the pseudoscalar gluonic operators.

\section{Acknowledgments}
Research funding from the Natural Science  \& Engineering Research Council
of Canada (NSERC) is gratefully acknowledged. Ailin Zhang is partly
supported by National Natural Science Foundation of China.

\end{document}